\documentclass[12pt,preprint]{aastex}
\shorttitle{}
\shortauthors{Nesvorn\'y et al.}
\begin{document}
\title{How to Find a Planet from Transit Variations}
\author{David Nesvorn\'y}
\affil{Department of Space Studies, Southwest Research Institute, 1050 Walnut St., \\Suite 300, 
Boulder, CO 80302, USA} 

\begin{abstract}
Here we describe a story behind the discovery of Kepler-46, which was the first exoplanetary system 
detected and characterized from a method known as the transit timing variations (TTVs).
The TTV method relies on the gravitational interaction between planets orbiting the same star. 
If transits of at least one of the planets are detected, precise measurements of its transit times 
can be used, at least in principle, to detect and characterize other non-transiting 
planets in the system. Kepler-46 was the first case for which this method was shown to work in practice. 
Other detections and characterizations followed (e.g., Kepler-88). The TTV method plays an important 
role in addressing the incompleteness of planetary systems detected from transits. 
\end{abstract}

My background is in dynamical astronomy. This is a branch of astronomy that is concerned with the motion of 
planets, stars and galaxies. Somehow, something clicked while I was attending lectures about 
dynamical astronomy as an undergraduate at the Charles University in Prague. I remember really enjoying the 
mathematical methods that we were asked to master. In other astronomy courses, we learned key equations with 
many variables. Their derivation was standard, they expressed interesting relations, but they left me unsatisfied. 
I think I really wanted to use advanced mathematics that we have just learned in other courses, and operating 
ratios of astrophysical variables was not exactly that. 

Now, the field of dynamical astronomy has a very long history, going back to Kepler, Newton, Laplace, Le Verrier,
Poincar\'e and Brouwer, to mention just a few names. The mathematical methods these guys developed are fantastic.
They include Hamiltonians, perturbation theories, resonances, chaos. Finally, I had the satisfactory feeling of 
actually using mathematics to understand fragments of the physical world. My colleagues got discouraged 
for exactly the same reason, and moved to other research, as far from the Hamiltonians as they could. The subject 
of my Master thesis was an obscure topic of orbital dynamics of asteroids near resonances with Jupiter, the scope 
of my doctorate was only inches broader. Still, these early days gave a good dose of useful scientific background.

The exoplanet discoveries in the past two decades were one of two revolutions that I witnessed at close range 
(the other one being the Velvet Revolution of 1989). New exoplanets are announced every month. The smallest, 
the one with the shortest orbital period, the most distant one, a system of three, four or even more planets. 
I remember watching these inspiring discoveries and wondering about how to make sense of the diversity 
that was emerging from observations. I still wonder about that now. 

The lack of data bothered me quite a bit in the early days, when exoplanet science was essentially an 
observational endeavor. Sure, one could start developing theories of this and that, but where is the assurance, 
given the paucity of constraints, that they could be correct. To this day, I still think that the detection and 
characterization of exoplanets is the top priority. 
That could be a dead end for a theorist like me, but back in 2007, after reading a few papers published a couple of 
years earlier, I got intrigued by the method known as the transit timing variations or TTVs for short. The method 
is straightforward. If there is just one transiting planet and no complications, the planetary transits must be occurring 
on a linear ephemeris. That is {\it exactly} on a linear ephemeris, as defined by the planet's orbital period.  
Once you add planets, however, they start pulling on each other, and their orbits are strictly Keplerian no 
more; compared to a linear ephemeris, some transits occur earlier and some later. That is what the TTVs stand for. 

The promise of the TTV method was that if there is, say, transiting Neptune closer in and an Earth-size planet 
farther out, then the inner planet's TTVs should reveal the existence of the outer planet. No other 
observations needed. Awesome, isn't it!? That was the theme that the first paper, by Holman \& Murray (2005), 
highlighted. The second paper, by Agol et al. (2005),  was broader in scope. It discussed TTVs for all sorts of planetary 
configurations and gave an approximate scaling of the TTV amplitude for each. 

So far so good, I recall thinking, but how about the heart of the matter, which is the {\it inverse} problem. The 
inverse problem arises when someone attempts to figure out the planetary parameters, such as the mass and orbit, 
from TTVs. Does the inverse problem have a unique solution? What if the companion planet is not transiting? 

My goal in 2007/2008 was somewhat naive, I thought it could be possible to fully resolve the inverse TTV problem by 
employing computer algebra. For that I considered a case where there are two planets orbiting the same 
star, and the orbital period ratio of the two planets is not in a ratio of small integers (i.e., the non-resonant 
orbits in a scientific jargon). There is a well known mathematical method, known as the Lie-Hori perturbation 
theory, that can be applied to this case. Furthermore, in practice, the observational coverage of any given system is 
relatively short, years to tens of years max. So, one only needs to consider the short-period 
variations, and ignore everything else. This turns out to be relatively easy. The interaction potential is expanded 
in the Fourier series and the short-period variations of orbital elements emerge in the Lie-Hori theory 
as derivatives of the generating function, which is closely related to the original Fourier expansion.
The variations are then plugged into an expression for TTVs, and voila, the TTVs are given as the 
Fourier series as well.

I reasoned that an observer can take the TTV data and perform the Fourier analysis, thus identifying frequencies, 
amplitudes, and phases of whatever terms the measurements contain. The short-period frequencies must be 
$k_1 n_1 + k_2 n_2$, where $k_1$ and $k_2$ are small integers, and $n_1$ and $n_2$ the orbital frequencies of the 
two planets (here, 1 is the inner transiting planet and 2 is the outer one). The periods of these terms are 
comparable to the orbital periods and that's why they are called the short-period TTVs. There is a slight complication 
as some identified frequencies can be aliases of $k_1 n_1 + k_2 n_2$. Still, given a set of frequencies, it should be 
possible to figure out which ones are real and which ones are aliases, eventually giving us $n_2$ (frequency 
$n_1$ is known from transits).

It then remains to match the observed amplitudes and phases to their theoretical expressions and we have a set of 
equations to compute unknowns, including masses of the two planets and their six orbital elements. That is, in 
general, 14 unknowns. Not all unknowns, however, can be determined from TTVs. For example, the unknown orbit orientation 
with respect to the sky plane causes a degeneracy in the nodal longitudes. The TTVs can be used to determine the 
difference, $\Omega_1-\Omega_2$, but not the individual values. Also, the true longitude of the transiting planet 
is known from the transit observations. So there are really only 12 unknowns. So, if the goal is to determine every 
parameter, at least 6 frequency terms need to be measured in the TTV signal. This gives six equations for phases 
and six for amplitudes, i.e., 12 equations and 12 unknowns, as it should be. 
 
Well, what seems simple in theory is not easy in practice, mainly because the equations are ugly and difficult to 
deal with. Rather than struggling with the analytic solution, I realized that it would more practical to do things 
numerically, and my goals shifted. Still, while my original plan somewhat predictably failed, I learned many 
things from this exercise. It occurred to me, for example, that the analytic method can be used to greatly speed up 
the whole process of inversion. 

To appreciate that, let's take a modern viewpoint on this issue and consider a purely numerical method. Say that 
an efficient N-body integrator is instructed how to compute the transit times for any given planetary system. The 
code is interfaced with some smart algorithm that knows how to maximize the likelihood of the fit. All that is OK, 
but the basic difficulty is that the algorithm must search in parameter space of 12 dimensions, which is a lot of 
dimensions.  The program may take too long to execute or it may not converge at all if the likelihood landscape 
is too complex. It all depends on how much CPU time it takes to compute transits for one planetary configuration.  

I developed a fully numerical code and tested it on mock planetary systems back in 2008. Things were annoyingly slow. 
So, I explored every avenue to speed things up. For example, do we gain anything if the N-body code is 
replaced by the Fourier routines that calculate TTVs analytically?  If so, they can perhaps be used to determine 
an approximate solution, and the N-body integrator can take over after that.

It turns out that the 12 variables can be split in four groups. The first and the most difficult to deal with
are the semimajor axes of the two planets. The coefficients of the Fourier series depend on them in non-trivial 
ways. Fortunately, they depend only on the ratio of the semimajor axis, $\alpha = a_1/a_2$, and not on $a_1$ and 
$a_2$ individually. So, it is possible to precompute all coefficients on a grid in $\alpha$ and devise a fast 
algorithm that interpolates from the grid to any value of $\alpha$. The likelihood of the fit is a very sensitive 
function of $\alpha$, so the grid must be sufficiently dense for things to work. [In fact, it is even better
to work with the orbital periods (rather than the semimajor axes), because of uncertainties in stellar mass.]   

The second group is the orbital eccentricities and inclinations. The coefficient dependence on them is simple:
they appear in all powers permitted by symmetries. It turns out to be possible to develop recursive routines to 
compute the higher powers from the lower ones such that the number of arithmetical operations is minimized.

The third group are the orbital angles, three for each planet. A really efficient way to evaluate the Fourier 
series for any combination of angles is to make use of complex algebra and symmetries. The algorithm computes
a few leading terms and combines them to get the remaining ones almost for free. Therefore, in essence, because 
the evaluation of the Fourier series for angles is so inexpensive, one can set aside the angle dimensions.
This effectively reduces the dimension of the problem.

The last group is the masses of the two planets. The TTVs of the inner transiting planet are nearly independent of 
its own mass and depend linearly on the mass of the outer planet. So, there is no hope, in absence of other 
information, to determine the mass of the transiting planet from its short-period TTVs. On the other hand, since 
the dependence on the perturbing planet's mass is linear, the algorithm can first compute TTVs for 
some indeterminate mass, and subsequently adjust $m_2$ to optimize the fit. Again, this is cheap. 

These are some of the main features of the code I developed in 2008.  In the final version, the algorithm based 
on the perturbation theory was {\it many orders of magnitude} faster than the N-body approach. It counts, you know,  
if the calculation can be done in minutes instead of many weeks. 

The radial velocity observations of warm and cold Jupiters indicate that these planets often have large 
orbital eccentricities. Their orbits were presumably excited by dynamical instabilities and gravitational scattering. 
So, to make my analytic algorithm applicable to these cases, I pushed the perturbation 
theory to very high powers of eccentricity, first to 5, then to 15, and finally to 25. This means that all 
terms in eccentricities up to power 25 were included in the final code. In retrospect, this was unnecessary 
because the {\it Kepler} observations showed that the TTV planets regularly have almost circular orbits. 

There were various hiccups while the code was being developed and tested. Some were more serious than others. 
In most cases, I was able to link these problematic cases to systems that were too compact, too close to 
resonances or something else, but sometimes a perfectly normal system was giving me a trouble. After weeks and 
weeks of struggle, with the problem seemingly going away and then coming back when I least expected it, I started 
to suspect that the Fourier expansion of the potential is at fault. I have adapted this part of the code from 
another program that was given to me by my former teacher and advisor, Milo\v{s} \v{S}idlichovsk\'y,
who developed and used it for other projects. 

It took some courage to start suspecting that something funky is going on in that part of the code, because 
Milo\v{s} is a very careful man. I would rather expect to find a bug in my code. Tracking the issue 
down, however, I found that some coefficients of the expansion are exactly two times smaller than they 
should be, and applied an empirical patch in the code that compensated for that. With that, things fell in 
place. I later visited Milo\v{s} just before his retirement from the Czech Academy of Sciences. When I described 
the problem to him, he said, of course, you are using the program to do things that it was not meant to do, 
and then proceeded to figure out what the real problem was. 

After 2008, I was ready to use my new algorithms to do things, but did not have any good data to try 
them on. Before {\it Kepler}, TTVs were measured for hot Jupiters and alike, which do not have planetary companions 
too often. Also, companions would need to be fairly close or near resonant period ratios for the TTV method to 
work. In addition, the ground-based transit observations produced TTVs with rather large measurement errors 
and sparse transit coverage. That was not good enough for solving the inverse problem. I knew that well because 
I was experimenting on mock systems, where I would inject realistic measurement errors in the TTV data and vary 
the number of observed transits. It turns out that even for an excellent signal-to-noise (Kepler-46 has 
$S/N \sim 50$, where $S/N$ is defined here as the ratio of the observed TTV amplitude to the timing 
measurement uncertainty), one still needs at least about 15 transits, continuous or not, for the inverse 
problem to have a unique solution.

The {\it Kepler} mission, of course, changed that, but not being part of the {\it Kepler} team did not help. Thus, I became
somewhat disheartened while watching the discoveries of Kepler-9b,c, which was the first planetary system characterized 
from TTVs (Holman et al. 2010), Kepler-19b, which was the first case that showed clear TTV evidence for an unseen 
planet (Ballard et al. 2011), and others. It looked as if the train departed leaving me behind. In the 
meanwhile, I improved the validity of the codes for strongly inclined systems and eccentric transiting 
planets, performed a tentative analysis of COROT-1b, TrES-1b and HD 189733b, and waited for something.

That something happened in November 2011 when I was on a sabbatical leave at the Nice observatory, in southern France.
I received an unexpected email from David Kipping, then Carl Sagan Fellow at the Harvard-Smithsonian Center for 
Astrophysics. David invited me to be a member of a small team of researchers known as the {\it Hunt for Exomoons with 
Kepler} (HEK; Kipping et al. 2012).\footnote{{\tt https://www.cfa.harvard.edu/HEK/}} Other then David and me, the 
HEK staff included 
G\'asp\'ar Bakos from Princeton University and Allan Schmitt from {\tt PlanetHunters.org}. Small teams suit me well 
and knowing David from before I gladly accepted. My role in the team was to use {\it Kepler} TTVs, which David extracted 
with formidable speed and accuracy from the MAST catalog, to characterize planets. 

The following part of the story, which eventually led to the detection and characterization of Kepler-46 and 
the publication of this work in the {\it Science} magazine, is best told by an email exchange among the HEK 
members. Below I reproduce excerpts from some of these emails along with short commentaries. In the following 
text, the dates of the emails are given in the US convention (MM/DD/YYYY).  

\noindent
{\bf From David K's email to the HEK team (11/27/2011):}

{\tt \footnotesize
HCV-439 is the most interesting system. The system exhibits TTVs 
of $\sim$1 hour amplitude whereas none of the other systems have `clean' 
TTVs like this. 

I attach here the TTVs of the top four candidates. HCV-439 is the only 
one with a really convincing signal. 

The highest priority is HCV-439. Could you please run your code on these 
TTVs and see what planetary solutions are valid (if any)? Let me know if 
you need any other system parameters. Note that the system is known to 
only have one transiting planet thus far. 
}

\noindent
Note that HCV-439, mentioned in the above email, was our HEK nickname for 
KOI-872, which later became known as Kepler-46. We used a nickname to avoid 
potential information leaks. 

\noindent
{\bf From David N's email on 12/1/2011:}

{\tt \footnotesize
I made some initial attempts to fit the TTV signal for 439. So far I only 
tried co-planar systems. There are many solutions that 
provide a good match to the data. 

The plot illustrates the importance of getting additional transits. With 
only one additional transit, it will be clearly possible to distinguish between 
the three solutions shown here. I expect that the unique solution will 
be found when we will have $\sim$10 transits.

This saw-tooth pattern looks to me like a short-period planetary signal 
which is ideal for my method (resonant oscillations would have longer period). 

One solution gives normalized chi2$<$0.1 and stands out as exceptionally good. 
This could be the right one. It corresponds to a planet half the Jupiter mass 
at $a=0.29$ AU and $e=0.1$. It may or may not be the right one.  

}

\noindent
It turns out that the best solution mentioned above is essentially the right one. 
This is amazing, because we only had 6 (!) transits of KOI-872b in late 2011,
when this email was written (Figure 1). 

\begin{figure}[t!]
\epsscale{0.4}
\plotone{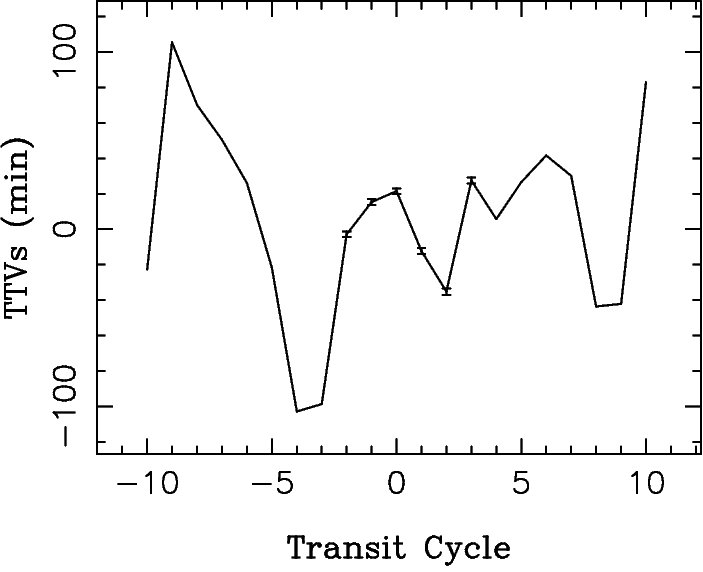}
\caption{From our early attempts to fit TTVs of KOI-872b.}
\label{fig1}
\end{figure}

\noindent
{\bf From David K's email on 12/1/2011:}

{\tt \footnotesize
Excellent stuff. It is frustrating that we only have 6 transits. In January we 
should acquire additional $\sim$10 transits so we will then have more than enough.

The TDVs are significant so I suspect there is important information in there 
and I look forward to seeing what happens when you fit the TTV and TDV 
simultaneously.

}

\noindent
David K was rightly pointing out the importance of transit duration 
variations (TDVs). The lack of TDVs of KOI-872 was later used to reject the 
second best solution and obtain a unique fit.

\noindent
{\bf From David N's email on 12/9/2011:}

{\tt \footnotesize
With 6 transits it is impossible to do anything reasonable with non-coplanar 
systems, so I focused on coplanar planetary fits.

For the moment I looked at the semimajor axis range of the putative planet between
0.17 to 1 AU. There are 30 solutions where analytic fits to observed TTVs showed
normalized chi2$<$1. This cut is arbitrary.

I would like to briefly highlight the first solution in the list below, 
corresponding to a planet about half the Jupiter mass with 80 day period and 0.1 
eccentricity. Attached plot shows the TTVs and TDVs for this system. TTVs look 
pretty good. 
}

\noindent
The `first' solution mentioned in my Dec. 9 email was published as the second-best solution 
(s2) in our {\it Science} article. This solution can be rejected with more TTV data and TDVs.

\noindent
{\bf From David K's email on 12/9/2011.}

{\tt \footnotesize 
This is all really cool. So either we have a moon or a 2nd planet then. I wonder 
if there is any way to constrain the mass of the transiting planet, even a broad limit 
such as Mp$<$10 Mj (i.e. a real planet rather than a false positive)? So what I'm 
really asking is can we confirm this candidate?
}

\noindent
Again, this is spot on. We could not obtain any limits on the mass of the transiting 
candidate from the TTV data available to us in 2011/2012. Instead, the candidate was 
confirmed by first detecting its companion from TTVs and then running a stability 
analysis for the whole system. This gives 6 Jupiter masses ($M_{\rm J}$) as an upper 
limit for KOI-872b, which is clearly planetary. 

In 2017, using the whole {\it Kepler} dataset and 35 transits of KOI-872b, the mass of the 
transiting candidate can already be constrained from TTVs to give 
$M_b=0.88_{-0.34}^{+0.37}$ $M_{\rm J}$ (Saad-Olivera et al. 2017). This is possible because the TTV signal 
starts picking up the non-linear terms in the mutual interaction of the two planets.

\noindent
{\bf From David K's email on 1/10/2012:}

{\tt \footnotesize
I want to just update on the progress of the real project here. I have adapted 
multinest to work on ``real'' data now, it took a hellish day of coding but I think it 
is now working. I am about to start a final test on HCV-439 to see if I recover the same 
TTVs as before. If it passes this test, then I will begin detrending the new {\it Kepler} data 
for this system. 

Once I have detrended the data, I will begin fits on all 14 transits. This could take 
some time, week or more. However, the nice thing with multinest is that one rarely has 
to re-execute the fits. 
}

\noindent
Multinest is a multimodal nested sampling algorithm that was developed by Farhan 
Ferroz (Ferroz et al. 2009). David K initially used it to obtain parameters of individual
transits from the {\it Kepler} photometry. We adapted the Multinest code to 
execute the dynamical fits as well (e.g., Nesvorn\'y et al. 2013).  

\noindent
{\bf From David K's email on 1/15/2012:}

{\tt \footnotesize
Cool, with such a complex and high SNR signal, I wonder if this could be the first 
example of a perturber being uniquely determined from TTV alone? Exciting stuff!

I am running refined TTV fits, they will take 1-2 days minimum.
}

\noindent
The next day David K sent me KOI-872's TTVs from the {\it Kepler} quarters 1-6, 
and I performed a more complete dynamical analysis immediately. Each dynamical fit was 
executed by the analytic codes in minutes (Figure 2). All the effort I have put into the code 
over the past three years finally paid off. Doing this with N-body is possible 
but takes days on a supercomputer.

\noindent
{\bf From David N's email on 1/16/2012:}

{\tt \footnotesize

It is looking good!

The analytic search routine in 5D (planet's mass,a,e,varpi,capm) found promising solutions
that fit TTVs very nicely (TDVs were ignored at this step).
Most of the solutions could not have been fine tuned with my numerical (and more) precise
code in 7D (mass,a,e,inc,capom,varpi,capm), so I discarded them. Two remained.

Solution 1: mass$=$0.000934 Msun, sema$=$0.2998 AU, ecc$=$0.0560              

This solution is related to the 2nd best solution that we obtained from the original 
6 transits. It is a Jupiter-mass planet just inside the 5:2 resonance (P2/P1$=$2.437) which explains 
large TTVs. The parameters are very well constrained, including the inclination. With inclination 
this large, TDVs can be substantial, but I did not have time to look into this in a more 
detail. Will do it asap.

Solution 2:  mass$=$0.0002107 Msun, sema$=$0.2346 AU, ecc$=$0.01224    

This is a planet 1/5 Jupiter mass near the 5:3 resonance (P2/P1=1.696). The ultimate 
test of this could be TDVs as this solution has smaller inclination than solution 1, 
and should produce smaller TDV amplitude. All this has to be checked with the new 
TDV data.

}

\begin{figure}[t!]
\epsscale{0.4}
\plotone{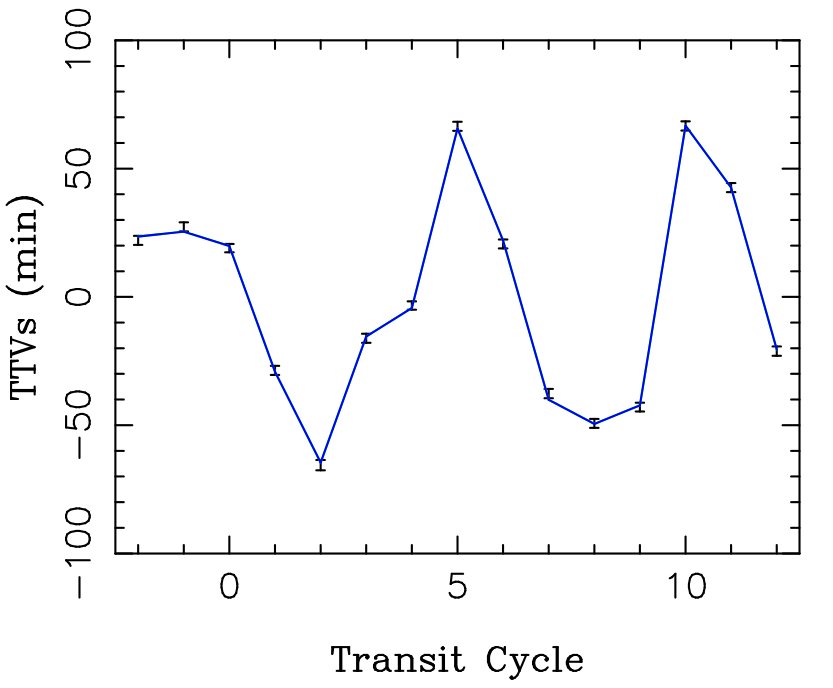}
\caption{A more definitive fit to TTVs of KOI-872b. We had 15 transits in total in January 
2012.}
\label{fig2}
\end{figure}

\noindent
Yes, this was a critical assessment: TDVs of the 2nd solution should be smaller. Given
that no TDVs were detected for KOI-872, the first solution can be ruled out because 
it generates TDVs in excess of the measurement errors.

By the way, my statement that ``the 5:2 resonance ... explains large TTVs'' is incorrect.
In fact, the large variations come from the 2:1 short-period term and are a consequence
of the relatively large companion mass.

\noindent
{\bf From David K's email on 1/16/2012:}

{\tt \footnotesize

Awesome! I have more good news. There are three epochs missing in the data I sent you. 
My code had a bug where it cut them off, but I have fixed it now. So you have 3 new 
transits to add in. 

}

\noindent
{\bf From David N's email on 1/17/2012:}

{\tt \footnotesize

Interesting news:
The priority of the two solutions that I mentioned in the previous email switched with 
the new data.
Now, the solution with a smaller planet near 5:3 gives chi2$=$5.8 (!) for 8 degrees 
of freedom. This one improved enormously.
The solution with the Jupiter-mass planet near 5:2 now gives chi2=28.9, and can be 
rejected.

I am starting to be confident that we have finally obtained the correct and unique 
solution.

To summarize:
In all likelihood we have here a system of two planets, the outer non-transiting planet 
should have mass about 25\% lower than Saturn. It should be near the 
5:3 resonance (period ratio 1.6967), low eccentricities and low inclinations. 

}

\noindent
We had 15 transits of KOI-872b and things were converging toward the solution near 
the 5:3 resonance, which is the correct one. And our excitement was building up...

\noindent
{\bf From David K's email on 1/17/2012:}

{\tt \footnotesize
Woah - that's a hell of a fit!

Transit wise - interesting point. I can begin looking for transits on the 5:3 resonance. 
My immediate instinct is that it cannot be transiting given the planet is likely at least 
Neptune radius and should be sticking out like a sore thumb. I should be able to produce 
constraints on inclination vs radius for a fixed period which will give you a strong 
constraint to add in.
}

\noindent
According to the best TTV/TDV fit mentioned above, the relative inclination of the two planets 
in the KOI-872 system must be small. We therefore considered the possibility that the outer 
planet might be transiting but was somehow overlooked. That turned out not to be the case:
no transits of the outer planet were detected.   

\noindent
{\bf From David N's email from 1/18/2012:}

{\tt \footnotesize
The short-period TTVs are sensitive to the Mc/M* ratio only, so I fit for Mc/M* 
and it comes out as $\sim$0.0002. For M*$=$0.8, this would roughly be a Saturn-mass planet.

It would be nice if transits of the 2nd planet can be ruled out. We could use it to give 
some lower limit on the perturbing planet's inclination.

Good news:
No interesting secondary maxima of chi2 popped out so far, so the two solutions that I 
mentioned in the previous email stand out as the only candidates. The 2nd solution can 
still be ruled out at >99\% confidence, leading to a unique parameter set. :-)
}

\noindent
With things going well, David K contacted the HEK team and told them the good news 
in the following email.

\noindent
{\bf From David K's email to the HEK team on 1/20/2012:}

{\tt \footnotesize
David N. and I have been working hard on HCV-439 these last few days are some 
answers are beginning to emerge. First of all, the new data strongly indicates the 
presence of star spots. Exomoon-like features seem to correlate with times of maximum 
activity which is a bad sign for exomoons. I have not run a moon fit through the new 
data yet, but my instincts are that this is not a moon. 

However, the TTVs persist and exhibit a complex and highly significant signal. 
David N. has managed to obtain what appears to be a unique TTV inverse fit to the data. 
Let me just stress that this is the 1st time this has ever been done by anyone for 
a non-transiting perturber and represents a major accomplishment if the solution 
holds with our subsequent tests. It seems as though an outer planet of about 0.25 
jupiter masses is near the 5:3 resonance and almost coaligned to the transiting 
planet. 

The near-coalignment suggests the outer planet has a good chance of transiting, 
but not guaranteed. Indeed, the fact it is a Neptune sized planet suggests it should 
have already been detected if it was there. I have run a search through the data and 
Allan has manually checked for transits but there is nothing convincing in the data.

So we are thinking of running a paper on this system and I am working hard on finishing 
up all of the fits for this system. Perhaps we should arrange a telecon next week 
sometime to discuss everything. Including journal, naming of the system, confirming 
the system, etc.
}

\noindent
It remained to confirm the planetary nature of KOI-872b. On the suggestion of David K,
I performed a stability analysis with different masses of KOI-872b. The results are 
described in the following email. 

\noindent
{\bf From David N's email to the HEK team on 1/30/2012:}

{\tt \footnotesize

I have done the stability test last week. The upper bound on Mb is 5 MJupiter for
M*$=$0.8 Msun, which is clearly planetary. I can increase M* and see what happens. 
The result can be predicted from the Wisdom's overlap criterion. Will update 
on this asap.

Please see attached a very preliminary draft of the paper that I sent to David K. 
already.

Given this result is so *cool* I think we should submit it to Science. I am now 
99.9\% sure that this is a real planetary system and that we identified the right 
solution. TTVs give us Mc/M*, Pc (or ac for assumed M*), ec$<$0.03, eb$<$0.02, 
ic$<$5 deg. We also have a very tight constraint on the pericenter and true longitudes.

Note that this is the first time that eccentricities were constrained from TTVs. The \\
inclination constraint is also unique. David K. just sent me new error estimates and 
things look even better that what is described in the attached draft.
}

\noindent
The stellar mass of KOI-872 was later revised to $0.90 \pm 0.04$ solar masses, indicating
that the mass of KOI-872b cannot be larger than about 6 Jupiter masses. 

We have done 17 iterations of the first draft and David K has written over 30 pages of the 
Supplementary Material. G\'asp\'ar Bakos, Lars Buchhave (Niels Bohr Institute) and Joel Hartman 
(Princeton University) improved the stellar parameters of KOI-872, and the whole HEK team 
was indispensable to the effort. Here it paid off how David K assembled the team with 
each member having a unique expertise. David was a driving force behind all efforts.
He also found transits of a third planet in the KOI-872 system, a super-Earth with 1.7 Earth 
radius and 6.8-day period. This planet should produce TTVs of KOI-872b of the order of 
seconds and has nothing to do with the measured TTVs.

I performed innumerous additional test to demonstrate that no other solution, including 
polar/retrograde orbits and other absurd configurations, and systems of multiple planets, 
can compete with the solution already found. Dozens of different fits were attempted overall.

\begin{figure}[t!]
\epsscale{0.7}
\plotone{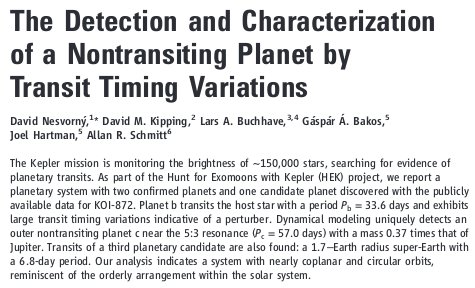}
\caption{Header of our {\it Science} article that announced the discovery on July 1, 2012.}
\label{fig3}
\end{figure}
  
The paper was submitted to {\it Science} on 2/26/2012. We received three referee reports on 
3/23/2012. This is from my email on 3/23/2012: 

{\tt \footnotesize

We received three *very* positive reviews!!! They are attached below.

Reviewer 1 praises the paper and has no criticism whatsoever. Referee 2 is similarly \\
enthusiastic and suggests only a few minor changes. Reviewer 3 liked the science as well 
(even congratulates us!) and offers numerous comments on how to improve the main text.

I have not seen such an uniformly positive reaction to any of the published Science/Nature 
papers that I contributed in the past. This is absolutely incredible and we should be proud 
of such an achievement. 

}

\noindent
The paper was finally published on June 1, 2012 (Figure 3), and David K and I prepared a press release 
to accompany the 
publication (Figure 4).\footnote{{\tt https://www.swri.org/press-release/unseen-planet-revealed-its-gravity}}
To reflect the confirmed nature of these planets, the {\it Kepler} team later renamed this system 
to Kepler-46. 

\begin{figure}[t!]
\epsscale{0.5}
\plotone{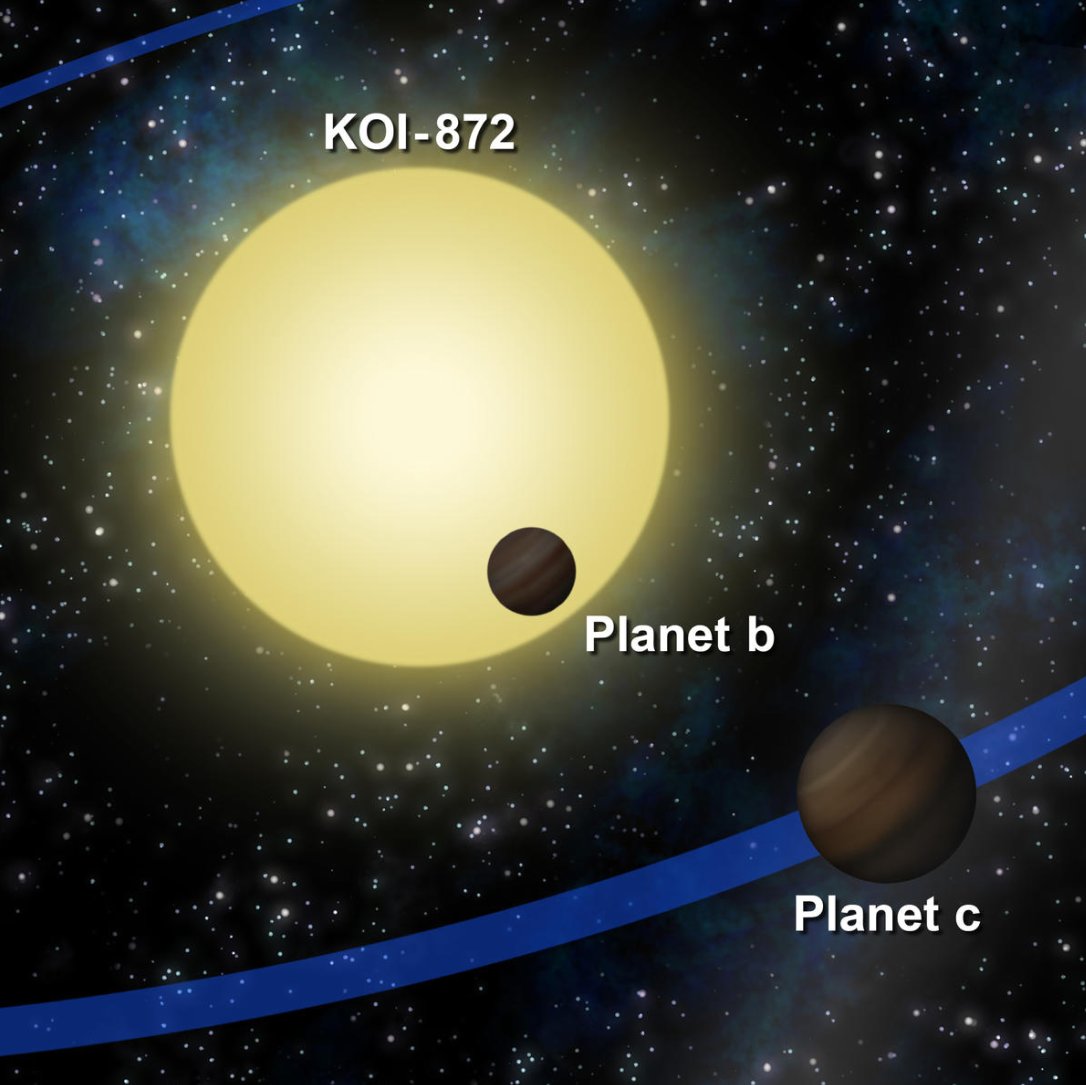}
\caption{An illustration from the SwRI press release.}
\label{fig4}
\end{figure}

What is, in retrospect, the importance of this work in planetary? To answer this question,
recall that TTVs were originally proposed as a non-transiting planet detection method, but prior to 
2012 they have found more use in validating the transiting planet candidates from {\it Kepler}. A 
non-transiting planet has been previously inferred via TTVs, but the measurement was unable to support 
a unique solution (Ballard et al. 2011). Our {\it Science} article demonstrated the full potential of 
TTVs as a method to detect non-transiting planets and precisely characterize their properties. So, 
in some sense, the work on Kepler-46 closed the loop and recasted the TTVs as a planet-detection 
method. 

This is useful because planetary transits can only be detected if the orbit is seen edge on.
The transit method is therefore blind to planetary companions with even a small orbital 
tilt away from the line of sight. The TTV method, on the other hand, can be used to detect  
non-transiting companions (assuming that at least one planet in the system is transiting). TTVs 
can therefore play an important role in addressing the incompleteness of planetary 
systems detected from transits, with interesting implications for the distribution of mutual 
inclinations of orbits. Ultimately, all this links to the holy grail of the exoplanet research, 
which is to establish how planets form and evolve.

How confident are we about the Kepler-46 system characterization published in 2012?
The full {\it Kepler} mission provided 35 transits of Kepler-46b (compared to 15 transits available 
to us back in 2012). My student, Ximena Saad-Olivera, recently performed a new TTV analysis with 
improved methods and all 35 transits (Saad-Olivera et al. 2017). This work confirmed that 
the Kepler-46c planet characterization, including its mass and orbital period, was correct. The
orbital eccentricities of Kepler-46b and c favored by the new fits are slightly higher than 
the original estimates, $e_{b,c}\simeq0.03$ versus $e_{b,c}\simeq0.01$-0.015, but
this is within the error bars reported in the original publication. Future TTV observations
including those of the Transiting Exoplanet Survey Satellite (TESS), can be used to further 
improve the Kepler-46 parameters.

Unfortunately, the host star of Kepler-46 is not bright enough (apparent magnitude 15.3) for 
precise Doppler observations. For this reason, it may take some time before Kepler-46c is confirmed 
by independent means, be it the radial velocity technique or something else. At this point, however, 
I think this is a mere formality, which brings us to a related story of KOI-142 (Kepler-88). 

We selected KOI-142 as an interesting TTV case from Mazeh et al. (2013), where they published a 
large collection of TTVs for dozens of {\it Kepler} candidates. The TTVs of KOI-142b are different
from those of Kepler-46b in that they show a huge, nearly-sinusoidal signal ($\sim$12 hour amplitude, 
which is about 5\% of the orbital period!). Also, we already had 105 transit epochs of KOI-142b 
in 2013.

I was initially not enthusiastic about this case, because KOI-142's TTVs did not seem to have the 
type of complexity required for the unique inversion. I was wrong. When David K provided a detailed 
analysis of transits and I attempted the dynamical fits in early 2013, the code very rapidly converged 
to a single solution --an outer planet just outside the 2:1 resonance with KOI-142b-- and nothing 
would move it from there.

This case is therefore unlike that of Kepler-46, where we only had a few transits to start with and 
were struggling with the solution ambiguity. Still, I did not understand why the solution should 
be unique in the case of KOI-142b until I realized that the measurements are so precise that 
they are picking up the chopping effect in the TTVs signal due to planet conjunctions 
(see Nesvorn\'y \& Vokrouhlick\'y 2014 for discussion of the conjunction effect). The chopping effect 
has a very low amplitude, of the order of minutes, and was buried in the huge TTVs produced by the 
near 2:1 commensurability between orbits. The chopping effect has a high information content and 
this is what was driving the code to converge to a unique solution.

Another happy moment with KOI-142 happened when we predicted TDVs from the best-fit TTV solution and 
then went back to the transit photometry to dig out the TDVs of KOI-142b. The measured TDVs turned 
out to be exactly where the dynamical solution of TTVs was predicting them. Such things do not 
happen by chance.

Our paper on KOI-142 was published in ApJ (Nesvorn\'y et al. 2013) and the system was later renamed to 
Kepler-88. Soon after, in 2014, new radial velocity measurements from the SOPHIE instrument were 
used to confirm the non-transiting planet Kepler-88c with the mass and orbital period that we
previously inferred from TTVs (Barros et al. 2014; the published SOPHIE velocimetry was not precise 
enough to improve the original parameter determination). This firmly demonstrated that the TTV 
method can be used to detect and characterize non-transiting planets, and resolved many doubts that 
I had back in 2007 when I started working on this project. More detections and characterizations 
of non-transiting planets followed later (e.g., KOI-227, KOI-319 and KOI-882; Nesvorn\'y et al. 2014).  

Kepler-88 is interesting because of its dynamical configuration near the 2:1 resonance. The 
orbital period ratio of the two planets is 2.03. This is therefore one of many pairs of the {\it Kepler} 
planets with orbits just outside of a first-order resonance, but in this specific 
case we have a very good determination of masses and orbital eccentricities. It is possible that 
Kepler-88b and 88c migrated into the resonance by gravitational torques from their parent gas disk and later 
separated by tidal migration. For that, however, the tidal dissipation would have to be unusually strong. 
Also, the orbital eccentricity of the outer planet, Kepler-88c, is substantial ($0.056 \pm 0.002$) and cannot 
be explained by gravitational perturbations from the inner planet. Perhaps there were, or still are, 
additional massive companions at larger orbital distances. In any case, the transits of Kepler-88b are 
predicted to disappear in 15-25 years from know (due to the precession of its orbital plane caused by 
Kepler-88c), so either these hypothetical outer companions reveal themselves in Kepler-88b's TTVs within 
the next two decades or we will have to use other methods (e.g., precise radial velocity measurements) 
to figure things out.         

\acknowledgements
Many thanks to the HEK team, and David Kipping in particular, for their work on the TTV-related issues.
The story of Kepler-46 and Kepler-88 described in this text would not happen without their vision, 
leadership and support. We thank the {\it Kepler} Science Team, especially the Data Analysis Working Group,
for making the {\it Kepler} data publicly available.  I would also like to thank Eric Agol, Katherine Deck, 
Daniel Fabrycky, Matt Holman, Alessandro Morbidelli, Jason Steffen and David Vokrouhlick\'y for numerous helpful 
discussions, and Jack Lissauer, Darin Ragozzine and Joann Ersberg for carefully reading the submitted 
manuscript and suggesting corrections.

\end{document}